# Interfacial Charge Transfer Driven Enhanced Transport and Thermal Stability in Graphene-MoS$_2$ Vertical Heterostructure Field-Effect Transistors

Ashis Kumar Panigrahi[1,2], Alok Kumar[1,2], Babulu Pradhan[1,2], Priyanka Sahu[1,4], Smruti Ranjan Senapaty[1,2], Monalisa Pradhan[3], Gopal K Pradhan[3], Satyaprakash Sahoo[1,2*]

[1]Laboratory for Low-Dimensional Materials, Institute of Physics, Bhubaneswar-751005, India

[2]Homi Bhabha National Institute, Training School Complex, Anushakti Nagar, Mumbai, 400094, India

[3]Department of Physics, School of Applied Sciences, KIIT Deemed to be University, Bhubaneswar, Odisha 751024, India

[4]School of Physics, Sambalpur University, Jyoti Bihar, Burla, Odisha 768019, India

[*]Corresponding Author E-mail: sahoo@iopb.res.in

**Abstract:**

Van der Waals (vdW) heterostructures developed by integrating two-dimensional (2D) materials offer a powerful route to engineer charge transport through interfacial effects. In this work, we demonstrate interfacial charge transfer driven transport enhancement in few-layer (FL) graphene-monolayer (ML) MoS$_2$ (Gr-MoS$_2$) vertical heterostructure field-effect transistor (FET). Raman scattering and Raman intensity mapping results confirm the successful stacking of FL graphene on ML MoS$_2$. Pronounced photoluminescence (PL) quenching of MoS$_2$ and spectral redshift in the heterostructure suggest efficient interlayer charge transfer and strong electronic coupling at the vdW interface. Electrical measurements show enhanced drain current, field-effect mobility, and conductivity in Gr-MoS$_2$ device compared to pristine MoS$_2$ transistor with Ag contacts. The energy band considerations under equilibrium and gate bias conditions suggest improved Fermi-level alignment and reduced effective Schottky barrier effects at the graphene-MoS$_2$ interface, enabling efficient carrier injection. Temperature-dependent transport (300-400 K) reveals phonon-dominated mobility and conductivity degradation in both devices; however, the heterostructure exhibits significantly suppressed performance degradation. The mobility enhancement factor increases from ~1.6 at 300 K to ~4.0 at 400 K, accompanied by a corresponding improvement in conductivity stability, demonstrating superior thermal robustness for the Gr-MoS$_2$ heterostructure. The power-law analysis indicates that transport in pristine MoS$_2$ is influenced by both intrinsic phonon scattering and additional thermally activated extrinsic processes such as contact and interfacial effects, whereas the weaker temperature dependence in the Gr-MoS$_2$ device reflects moderated extrinsic contributions and transport behaviour approaching a predominantly phonon-limited regime. These findings demonstrate graphene contact engineering as a viable pathway toward improved performance and thermally stable two-dimensional semiconductor electronics.

**Keywords:** Gr-$MoS_2$, FET, mobility, Schottky barrier, carrier transport

**Introduction**

Two-dimensional (2D) materials have emerged as a vibrant platform for next-generation electronic and optoelectronic technologies owing to their exceptional physical, optical, and electronic properties that are distinctly different from their bulk counterparts.[1,2] Among them, graphene[3] and transition metal dichalcogenides (TMDs),[4] particularly molybdenum disulfide ($MoS_2$),[5] have attracted tremendous attention due to their complementary characteristics and potential for heterogeneous integration.[6,7] Graphene (Gr), composed of carbon atoms arranged in a planar honeycomb lattice, exhibits extraordinary carrier mobility reaching values as high as ~200,000 $cm^2$ $V^{-1}$ $s^{-1}$ at room temperature, making it an appealing candidate for high speed electronic applications.[3,8–10] However, the absence of an intrinsic bandgap in graphene leads to poor current switching behaviour, which severely limits its standalone use in logic and digital electronics, despite extensive efforts such as chemical functionalization, doping, graphene nanoribbon fabrication, and bilayer graphene with dual-gate architectures.[11,12] In contrast, monolayer (ML) $MoS_2$ is a direct bandgap semiconductor with strong spin-orbit coupling and robust photoluminescence (PL), rendering it highly suitable for optoelectronic, valleytronic, and sensing applications.[13,14] Structurally, hexagonal $MoS_2$ consists of a ML of molybdenum atoms sandwiched between two layers of sulfur atoms and typically exhibits intrinsic n-type semiconducting behaviour. $MoS_2$-based field-effect transistors (FETs) are known for their high ON/OFF current ratios, often exceeding $10^7$-$10^8$, which is critical for low-power and high-contrast switching operations.[15,16] Despite these advantages, the intrinsic carrier mobility of CVD grown $MoS_2$ remains relatively modest, typically in the range of 0.1-10 $cm^2$ $V^{-1}$ $s^{-1}$.[17] This value is several orders of magnitude lower than that of graphene and imposes a major bottleneck for the realisation of high-speed and high-performance $MoS_2$-based electronic devices.

To overcome these limitations, extensive efforts have been devoted to enhancing the electrical transport properties of MoS$_2$ through dielectric engineering, substrate modification, and optimisation of contact resistance.[18–21] In particular, the use of high-κ dielectric materials such as Al$_2$O$_3$ and HfO$_2$ has proven effective in improving dielectric screening, suppressing Coulomb scattering, and mitigating surface reactions, leading to mobility enhancement up to ~80 cm$^2$ V$^{-1}$ s$^{-1}$ in optimised devices.[22–24] While these approaches have yielded significant improvements, they often involve complex fabrication steps, limited scalability, or do not simultaneously address contact-related transport bottlenecks.

The ability to vertically stack dissimilar 2D materials with atomic precision has opened new avenues for engineering artificial vdW heterostructures with tailored electronic properties.[6,25,26] In this context, heterostructures composed of few-layer (FL) graphene and ML MoS$_2$ offer a compelling strategy to synergistically combine the high mobility of graphene with the excellent switching characteristics of MoS$_2$.[27] The atomically sharp interface, absence of dangling bonds, and weak interlayer vdW interactions enable efficient charge transfer, reduced interface trap density, and improved contact quality. Furthermore, the integration of graphene as a contact material provides a versatile platform for engineering the metal-semiconductor interface, enabling tunable charge injection and improved control over contact resistance, which is crucial for optimizing device performance.[28] In recent times, there have been a great research interest for all 2D TMDC FETs with graphene contacts as source and drain, the preliminary results are very promising for minimizing device architecture as well as for future integration in CMOS technologies.[29] It is also interesting to investigate the carrier transport in TMDC FETs having asymmetric drain source contact especially a TMDC FET configuration having source as metal and drain as graphene contact. To the best of our knowledge such TMDC FET architecture has not been investigated so far. A deliberate tuning of interfacial charge transfer and its impact on temperature-dependent transport in such architecture could bring more insights into transport studies. Additionally, a systematic investigation is essential to understand how Schottky barrier modulation, phonon scattering, and interfacial trap states together influence carrier transport at higher temperatures thus warranting further studies.

In this work, we report interfacial charge transfer driven transport enhancement in FL graphene-ML MoS$_2$ (Gr-MoS$_2$) vertical heterostructure field-effect transistors based on scalable CVD-grown ML MoS$_2$. Structural and optical characterisations (Raman mapping, Raman and PL spectroscopy, AFM analysis) confirm the formation of a clean FL Gr-ML MoS$_2$ interface, with ML thickness (Δω ≈ 19.7 cm$^{-1}$) and pronounced PL quenching suggesting strong interlayer coupling and efficient charge transfer. Band alignment considerations under equilibrium and gate bias conditions suggest Schottky barrier modulation and gate-tunable carrier redistribution governed by graphene's work function and Fermi-level alignment. Electrically, the heterostructure devices exhibit enhanced drain current, field-effect mobility, and conductivity compared to pristine MoS$_2$ transistors with Ag contacts. Temperature-dependent transport (300-400 K) demonstrates phonon-dominated behaviour in both systems; however, the Gr-MoS$_2$ devices show significantly suppressed mobility degradation, with the enhancement factor increasing up to ~4 at 400 K, confirming indicating improved thermal stability of carrier transport. Power-law fitting ($\mu, \sigma \propto T^{-\gamma}$) yields exponents of γ ~ 5.124 and 5.246 for mobility and conductivity in pristine MoS$_2$, whereas reduced values of γ ~ 1.985 and 2.657 are observed for the Gr-MoS$_2$ device. This corresponds to a reduction factor of ~2.6 for mobility and ~2.0 for conductivity, indicating reduced temperature sensitivity of carrier transport. The improved carrier injection and atomically sharp vdW interface in the Gr-MoS$_2$ architecture are likely responsible for mitigating temperature-activated extrinsic effects, thereby enabling transport behaviour that approaches a predominantly phonon-limited regime. Collectively, these results demonstrate that graphene-contact engineering as a scalable and effective strategy for achieving high-performance, thermally stable 2D electronic platforms.

**Experimental Methods:**

**Synthesis of Monolayer MoS$_2$**

Chemical vapour deposition (CVD) technique is employed for the synthesis of monolayer (ML) MoS$_2$ on SiO$_2$/Si substrates. High-purity molybdenum trioxide (MoO$_3$, 1 mg, 99.9%) and sulfur powder (S, 100 mg, 99.9995%) are used as the metal and chalcogen precursors, respectively. The precursors are placed in separate alumina boats inside a one-zone horizontal tube furnace, with their relative positions adjusted along the furnace temperature gradient. Si substrates (1 × 1 cm$^2$)/SiO$_2$ (300 nm) are positioned face-down above the MoO$_3$ precursor boat to promote uniform nucleation and growth. During the growth process, carrier gas (Ar) is introduced at a flow rate of 50 sccm to transport the sulfur vapour toward the MoO$_3$ source and facilitate the sulfidation reaction. The distance between the $MoO_3$ and sulfur precursor boats is optimised such that sulfur vaporisation occurred at approximately 200°C, coinciding with the onset of $MoO_3$ evaporation at around 750°C. The furnace temperature is maintained at 750°C for 15 minutes, allowing the formation of ML and layered $MoS_2$. After completion of the growth process, the furnace is naturally cooled to room temperature, resulting in the formation of layered $MoS_2$ crystals with lateral dimensions of a few micrometres on the $SiO_2$/Si substrates.

**Preparation and Transfer of Graphene onto $MoS_2$**

Few-layer (FL) graphene flakes are mechanically exfoliated from natural graphite using the conventional scotch tape method and transferred onto a $SiO_2$/Si substrate. Selected graphene flakes are then transferred onto the as-grown ML $MoS_2$ using a polymethyl methacrylate (PMMA) assisted dry transfer method, following a procedure reported elsewhere.[30] The PMMA supported graphene is carefully aligned and placed onto the $MoS_2$ surface under an optical microscope, followed by PMMA removal using acetone, yielding a clean Gr-$MoS_2$ heterostructure.

**Device Fabrication**

Three terminal devices based on ML $MoS_2$ and Gr-$MoS_2$ heterostructure are fabricated using photolithography with a Heidelberg μPG101 pattern generator. After graphene transfer, the CVD grown $MoS_2$ samples on a $SiO_2$/Si substrate are first coated with a positive photoresist (ma-p-1201) using a spin coater (SUSS Microtech), followed by soft baking at 80°C for 1 min. Under optical microscope inspection, contact patterns are exposed onto selected FL Gr-ML $MoS_2$ heterostructure regions and ML $MoS_2$ flakes using a 405 nm laser source. The exposed patterns are developed in an alkaline solution (NaOH/DI water, 1:4) for 1 min, and then they are baked for 1 min at 80°C, after which the samples are transferred to a thermal evaporation chamber for silver (Ag) electrode deposition. Subsequently, the lift-off process is carried out by dissolving the residual photoresist in acetone for 20 min, yielding well-defined source and drain contacts. The heavily p-doped silicon substrate served as the global back-gate electrode, while the 300 nm thick $SiO_2$ layer functioned as the gate dielectric.

**Electrical Characterization:**

Electrical characterization of the ML $MoS_2$ and Gr-$MoS_2$ heterostructure based devices are carried out using a Keithley 4200A-SCS semiconductor parameter analyzer. The fabricated devices are mounted on a cryogenic four-probe probe station (Lake Shore). All measurements are conducted under high-vacuum conditions (~$10^{-6}$ mbar) to suppress environmental effects such as moisture adsorption and oxygen-induced charge trapping. To eliminate any contribution from photo-generated carriers, all electrical measurements are performed in dark conditions. Temperature-dependent output and transfer characteristics are recorded over a wide temperature range to examine the evolution of device hysteresis, threshold voltage shifts, and mobility modulation.

**Raman and PL Spectroscopy**

Raman spectroscopy and photoluminescence (PL) measurements are performed using a confocal micro-Raman spectrometer (Renishaw InVia) equipped with a 532 nm laser excitation source in a backscattering geometry. A 100× objective lens (NA = 0.8) is employed to achieve high spatial resolution. For both PL & Raman measurements, laser power and acquisition parameters are optimised to prevent any laser-induced local heating, sample degradation and thermal drift while ensuring a high signal-to-noise ratio. These optical characterizations are employed to confirm the monolayer nature, crystalline quality, and spatial uniformity of the synthesized $MoS_2$ flakes, as well as to verify the successful formation and interfacial integrity of the graphene-$MoS_2$ heterostructure.

**Results and Discussions:**

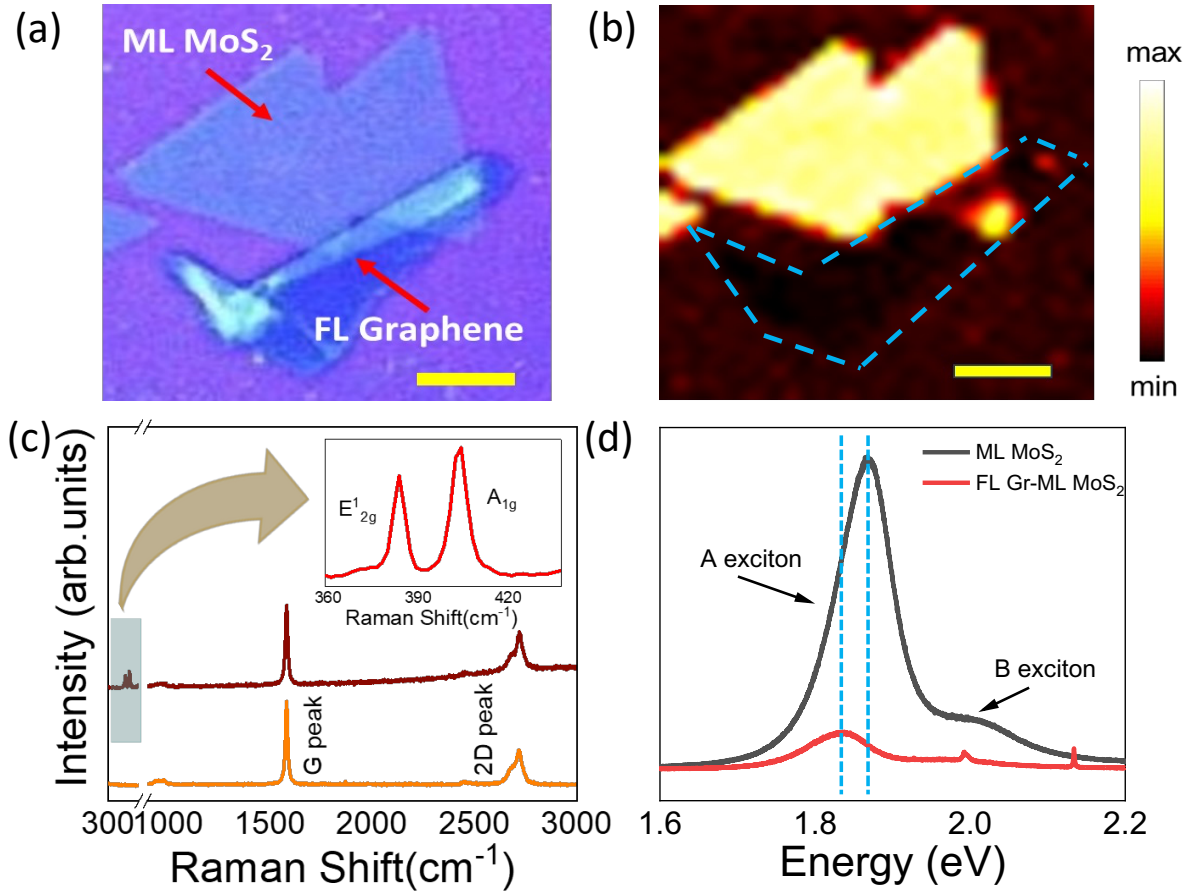

**Figure 1:** *(a) Optical image of the few-layer (FL) graphene- monolayer (ML) MoS$_2$ (Gr-MoS$_2$) vertical heterostructure. (b) Raman intensity mapping corresponding to the A$_{1g}$ peak of ML MoS$_2$. The dotted region corresponds to FL graphene flake. Scale bar 10μm. (c) Comparison of Raman spectra for FL graphene, FL Gr-ML MoS$_2$ heterostructure. The inset shows the characteristic Raman modes of ML MoS$_2$ observed in the heterostructure spectrum. (d) Comparison of PL spectra for ML MoS$_2$ and Gr-MoS$_2$ heterostructure. The dotted lines highlight a distinct red shift observed in the FL Gr-ML MoS$_2$ heterostructure spectrum.*

Figure 1a presents the optical microscopic image of the fabricated Gr-MoS$_2$ heterostructure, where a few-layer (FL) graphene flake is mechanically exfoliated and subsequently transferred onto a monolayer (ML) MoS$_2$ triangular domain using a dry transfer technique with a PDMS stamp. The optical images of ML MoS$_2$, FL graphene and Gr-MoS$_2$ heterostructure are shown in Fig. S1 of the supplementary information. The contrast difference between the two materials clearly distinguishes the underlying MoS$_2$ and the overlaid graphene regions, confirming a clean and well-aligned transfer. Figure 1b and Fig. S2(a) (supplementary information) show the Raman intensity mappings corresponding to the A$_{1g}$ vibrational mode of MoS$_2$ and the G mode of graphene, respectively. The bright regions in the A$_{1g}$ map (Fig. 1b) highlight the spatial distribution of MoS$_2$, while those in the G-mode map (Fig. S2(a)) correspond to the graphene region, confirming the successful formation of a vertically stacked FL Gr-ML MoS$_2$ heterostructure. The comparative Raman spectra, shown in Fig. 1c, are recorded from three representative regions: FL graphene, ML MoS$_2$ (shown in the supplementary information, Fig. S3(b)) and FL Gr-ML MoS$_2$ heterostructure. The ML MoS$_2$ exhibits the characteristic E$^1_{2g}$ (in plane vibration) and A$_{1g}$ (out of plane vibration) phonon modes located at 383.4 cm$^{-1}$ and 403.1 cm$^{-1}$, respectively, giving a frequency separation (Δω ~19.7 cm$^{-1}$) consistent with the ML nature of MoS$_2$. The Raman spectrum of graphene shows prominent peaks at 1583 cm$^{-1}$ (G band) and 2714 cm$^{-1}$ (2D band), confirming its FL nature. Individual Raman spectra for FL graphene and ML MoS$_2$ are shown in Fig.S3 (a, b) respectively. The observation of these Raman signatures confirms the presence and quality of both materials in the heterostructure. The photoluminescence (PL) spectra, presented in Fig. 1d, further demonstrate the electronic interaction between the graphene and MoS$_2$ layers. Pristine MoS$_2$ exhibits a strong direct excitonic emission centered at 1.83 eV, characteristic of ML MoS$_2$. It is very interesting to note that, the PL spectra of the underneath MoS$_2$ recorded on the graphene surface shows significant quenching in the PL intensity, along with a red shift, indicating efficient interlayer charge transfer from MoS$_2$ to graphene. The PL quenching in the Gr-MoS$_2$ heterostructure is attributed to efficient interfacial charge transfer from MoS$_2$ to graphene, which suppresses excitonic radiative recombination. Owing to the zero bandgap and high carrier mobility of FL graphene, photogenerated carriers in MoS$_2$ are rapidly transferred to the graphene layer prior to recombination, resulting in reduced PL intensity. The accompanying red shift suggests band-alignment-induced modification of the excitonic transition due to interfacial charge redistribution.[31] Consistent with our findings, Hwang et al. also reported pronounced photoluminescence quenching in graphene-MoSe$_2$ heterostructures, attributed to efficient interlayer exciton and charge transfer, thereby emphasizing the crucial role of interfacial coupling in governing the optical response of such vdW heterostructures.[32] Atomic force microscopy (AFM) topography of the Gr-MoS$_2$ heterostructure shown in supplementary information Fig.S4 (a, b) reveals well-defined triangular MoS$_2$ domains. The corresponding height profile shows a thickness of ~5.2 nm, corresponding to approximately 15 graphene layers, assuming an interlayer spacing of ~0.34 nm. These observations collectively confirm the overall architecture as a FL Gr-ML MoS$_2$ vertical heterostructure.

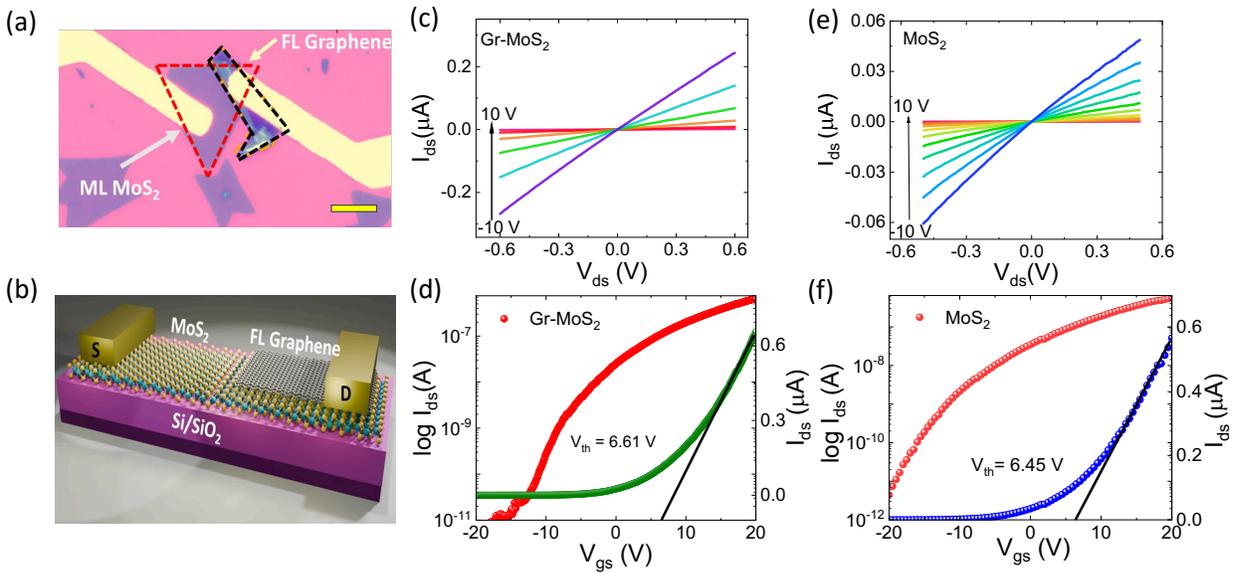

Figure 2: *(a) Optical image of the fabricated device based on the Gr-MoS$_2$ heterostructure. The MoS$_2$ regions are outlined by red dotted lines, while the graphene regions are indicated by black dotted lines. Scale bar 10μm. (b) Schematic illustration of the fabricated Gr-MoS$_2$ device. Output characteristics ($I_{ds}$-$V_{ds}$) and transfer characteristics ($I_{ds}$-$V_{gs}$) shown in both linear and logarithmic scales of device based on (c, d) Gr-MoS$_2$ heterostructure and (e, f) pristine ML MoS$_2$ respectively.*

Figure 2a presents the optical micrograph of the fabricated back-gated field-effect transistor (FET) device based on Gr-MoS$_2$ heterostructure. In the Gr-MoS$_2$ heterostructure device, electrical contacts are designed such that the source (Ag metal contact) is directly deposited on the exposed MoS$_2$ region. For electrical measurement purposes, Ag electrode is also carefully deposited on the graphene flake. The graphene contact on MoS$_2$ acts as drain electrode. Consequently, the Ag electrode on graphene does not make direct contact with the MoS$_2$ channel but interfaces with it through the graphene layer. This configuration results in an asymmetric contact geometry, comprising a direct metal-semiconductor (Ag-MoS$_2$) contact on one side and a graphene-mediated contact (Ag-graphene-MoS$_2$) on the other. Figure 2b shows the schematic representation of the device geometry. The output characteristics ($I_{ds}$-$V_{ds}$) and the transfer characteristics ($I_{ds}$-$V_{gs}$) of the Gr-MoS$_2$ device at room temperature (RT) are presented in Figs. 2c and 2d, respectively. For the sake of completeness, the output and transfer characteristics of MoS$_2$ transistor without graphene contact are also carried out which are shown in Figs. 2e and 2f respectively. The optical microscopic image of the device fabricated over MoS$_2$ is shown in the supplementary information Fig. S5. By analysing the output characteristics of both devices (Fig. 2c and 2e), it is evident that they exhibit linear current-voltage behavior at low drain bias, suggesting near-ohmic contact behaviour. The drain current ($I_{ds}$) lies in the microampere (µA) range, demonstrating the high crystalline quality and low defect density of the CVD grown MoS$_2$ channels. The $I_{ds}$ increases systematically with gate voltage ($V_{gs}$), reflecting effective gate electrostatic control over the channel conductance. Electrical transport measurements are carried out under high vacuum (~$10^{-5}$ mbar) to minimise environmental effects such as adsorbate induced doping or scattering. The transfer characteristics ($I_{ds}$-$V_{gs}$) at a constant $V_{ds}$ = 1V for both devices are shown in Fig. 2d and 2f. The $I_{ds}$ increases significantly with positive gate voltage, confirming typical n-type semiconductor behaviour for both the heterostructure as well as MoS$_2$ devices. Additionally, the transfer characteristics at different drain voltages $V_{ds}$ = (0.05-1 V) are presented in Fig. S6(a, b) of the supplementary information for the Gr-MoS$_2$ heterostructure device and pristine ML MoS$_2$ device, respectively. In both cases, the drain current increases with gate voltage. Notably, the Gr-MoS$_2$ device exhibits significantly higher $I_{ds}$ compared to pristine MoS$_2$, indicating improved carrier injection at the contact interface. The systematic increase in $I_{ds}$ with increasing $V_{ds}$ further reflects enhanced carrier transport and strong electrostatic modulation. This improvement can be attributed to improved carrier injection and efficient charge transfer at the graphene-MoS$_2$ interface. Leong et.al., in a separate study, demonstrated that mono and bilayer graphene contacts enable reduced contact resistance, thereby improving carrier transport in MoS$_2$ FETs.[28] A detailed explanation of our observed results will be discussed later in the proposed mechanism section.

The strong gate modulation from −20 V to +20 V demonstrates efficient electrostatic control of the channel. The field-effect mobility ($\mu_{FE}$) is extracted using the standard relation:

$$\mu_{FE} = \frac{dI_{ds}}{dV_{bg}} \times \frac{L}{W C_{ox} V_{ds}} \qquad (1)$$

where L and W are the channel length and width, and $C_{ox}$ is the gate capacitance per unit area of the 300 nm SiO$_2$ dielectric layer, calculated as $C_{ox} = \frac{\varepsilon_0 \varepsilon_r}{d}$ with $\varepsilon_r$ = 3.9. Using L = 3.85 μm, W = 6.92 μm for MoS$_2$ and L = 4.2 μm, W = 8.2 μm for Gr-MoS$_2$, the extracted mobilities are found to be approximately 3.84 cm$^2$V$^{-1}$s$^{-1}$ for MoS$_2$ and 6.23 cm$^2$V$^{-1}$s$^{-1}$ for Gr-MoS$_2$ at room temperature indicating an enhancement factor of ~1.62 in the presence of graphene. A similar trend has been reported by Hoque et al., where MoS$_2$ FETs with conventional metal contacts exhibited relatively low mobilities (~1.9 cm$^2$V$^{-1}$s$^{-1}$), while the incorporation of graphene contacts significantly improved the mobility to ~14.5 cm$^2$V$^{-1}$s$^{-1}$.[29] The threshold voltage ($V_{th}$) is extracted from the linear extrapolation of the transfer curves ($I_{ds}$-$V_{gs}$) curves in the linear regime, yielding $V_{th}$ ~ 6.45 V for MoS$_2$ and 6.61 V for Gr-MoS$_2$. Using these calculated mobilities and the gate-induced carrier densities, the conductivity of the devices is estimated from,

$$\text{Conductivity}, \sigma = \mu.q.n = \mu.q.\frac{C_{ox}(V_{gs}-V_{th})}{q} = \mu.C_{ox}(V_{gs}-V_{th}) \quad (2)$$

Where n is the carrier density $(n = \frac{C_{ox}(V_{gs}-V_{th})}{q})$.[17,33] The conductivities are calculated to be 3.45 × 10$^{-7}$ S/m for MoS$_2$ and 4.94× 10$^{-7}$ S/m for Gr-MoS$_2$ for their respective gate overdrive voltages (14.28 V and 13.51 V for MoS$_2$ and Gr-MoS$_2$, respectively). The observed improvement in drain current, mobility and conductivity for Gr-MoS$_2$ device at RT is attributed to the atomically sharp and defect-free graphene-MoS$_2$ interface, which minimizes Fermi-level pinning, reducing contact resistance thereby enabling efficient interfacial charge transfer.[29] This is expected to reduce the effective Schottky barrier at the Gr-MoS$_2$ junction, where graphene, because of its high carrier mobility, finite density of states near the Fermi level under operating conditions, acts as a low-resistance conductive pathway that promotes charge delocalization, screens Coulomb impurities, and suppresses interfacial trap states.[28,29] These results are consistent with the PL quenching behaviour discussed earlier, supporting strong electronic coupling between the layers. Overall, these observations demonstrate that graphene coupling effectively enhances the electrical transport properties in MoS$_2$ channels at room temperature (RT).

**Possible Mechanism:**

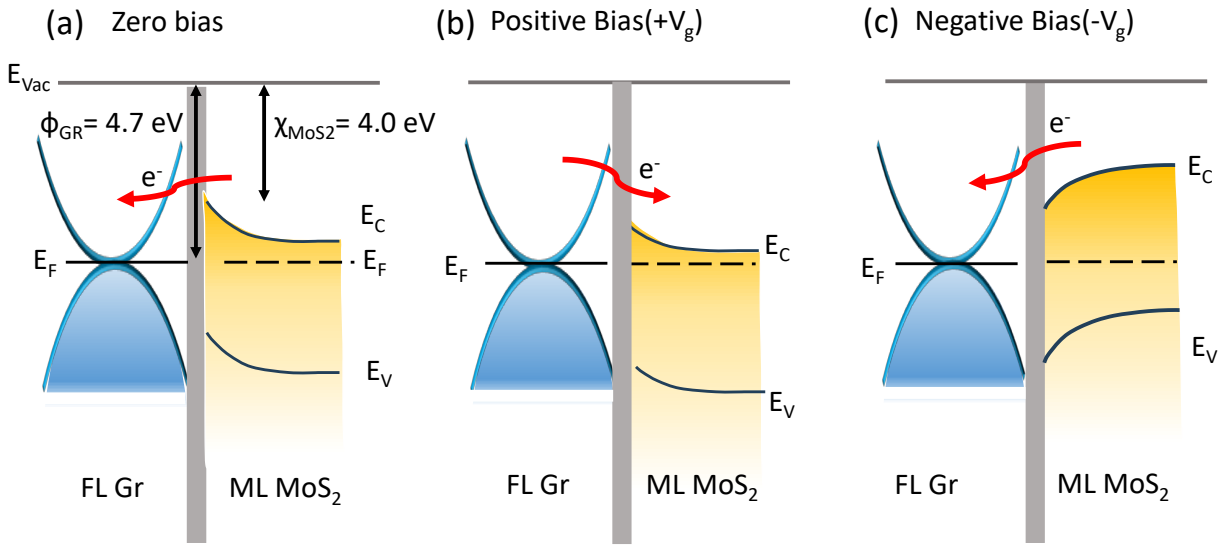

***Figure 3:*** *Energy band diagrams of the Gr-MoS$_2$ heterostructure with Ag back gate under (a) equilibrium ($V_g = 0$), (b) positive bias ($+V_g$), and (c) negative bias ($-V_g$).*

A detailed discussion of the enhanced transport properties in the Gr-MoS$_2$ devices is presented below. At RT, the Gr-MoS$_2$ heterostructure device clearly exhibits a significantly higher drain current along with enhanced mobility and conductivity compared to pristine MoS$_2$ device (Fig. 2). To understand the physical origin of this enhancement, we propose a possible mechanism based on interfacial band alignment and charge transfer, as schematically illustrated in Fig. 3. In the designed heterostructure comprising FL graphene and ML MoS$_2$, with silver (Ag) serving as both the back gate and electrode, the charge transfer and band alignment are governed by the relative work functions and the applied gate bias. At equilibrium (zero bias), the Fermi levels of graphene and MoS$_2$ align. Owing to the difference between the work function of FL graphene(~4.7 eV),[34] and the electron affinity of MoS$_2$ (~4.0 eV), charge redistribution occurs at the interface, leading to electron transfer from MoS$_2$ to graphene and the establishment of a built-in electric field directed from MoS$_2$ toward graphene.[35] When a positive gate bias ($+V_g$) is applied to the Ag back gate, the induced electric field attracts electrons toward the MoS$_2$ channel, causing downward band bending and reducing the effective Schottky barrier height for electron injection from graphene. Consequently, electron concentration in MoS$_2$ increases, placing the device in an accumulation regime characterized by enhanced n-type conduction and improved charge injection. At RT, this interfacial charge transfer lead to a significant enhancement in field-effect mobility and conductivity in the Gr-MoS$_2$ device compared to pristine MoS$_2$, supporting the role of graphene as an effective conductive bridge that facilitates carrier injection.[36] Conversely, under a negative gate bias (-V$_g$), upward band bending occurs in MoS$_2$, increasing the effective Schottky barrier and suppressing electron injection. The electron density in the MoS$_2$ channel is significantly reduced, driving the device into a deep depletion regime. Importantly, MoS$_2$ remains n-type, thus no hole inversion occurs. Instead, the reduced carrier density and weakened interfacial charge transfer lead to a decrease in conductance. Thus, the applied gate bias effectively modulates band bending, interfacial charge transfer, and carrier concentration in the Gr-MoS$_2$ heterostructure.

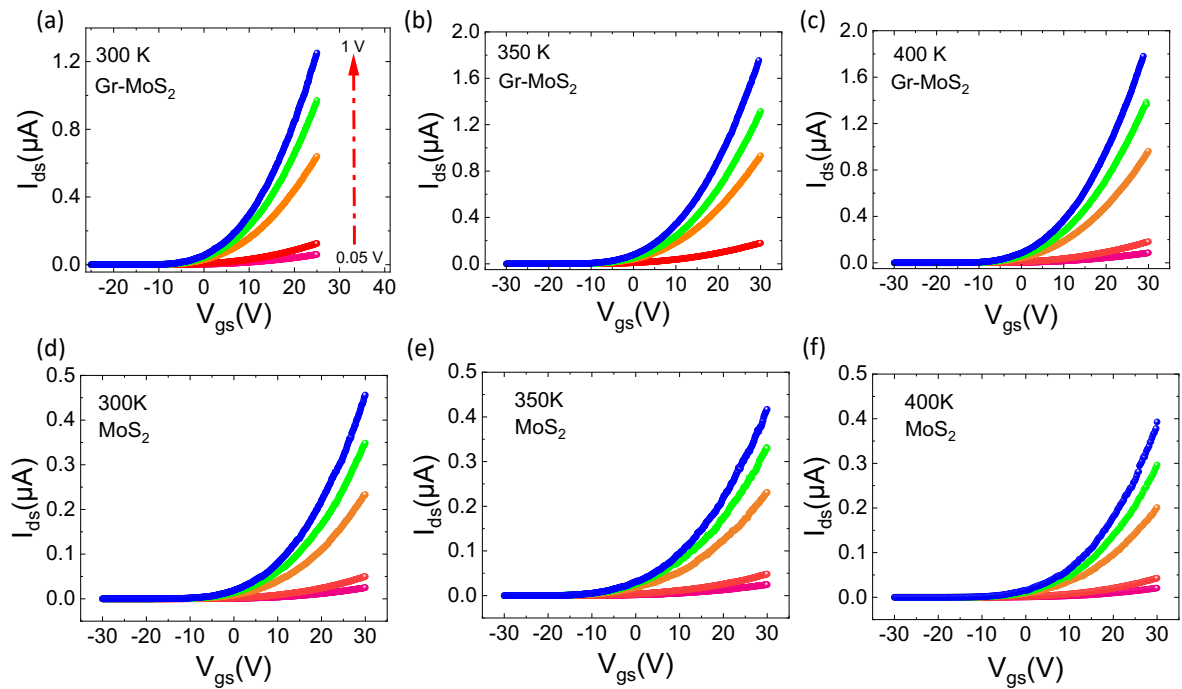

*Figure 4:* *Temperature-dependent transfer characteristics (I$_{ds}$-V$_{gs}$) of devices fabricated over (a-c) Gr-MoS$_2$ heterostructure and (d-f) ML MoS$_2$ field-effect transistors measured at drain voltages ranging from V$_{ds}$=0.05 to 1.0 V (shown in red dotted line). The measurements are systematically performed at temperatures of 300 K (room temperature), 350 K and 400 K, respectively.*

The above discussion emphasizes that interfacial charge transfer plays a crucial role in enhancing the drain current, mobility, and conductivity of the Gr-MoS$_2$ heterostructure FET at RT. However, it is equally important to examine how these interfacial effects evolve with temperature and influence the transport behaviour under temperature-dependent conditions. Figure 4(a-f) and Fig. S7(a-f) present the temperature dependent transfer characteristics (I$_{ds}$-V$_{gs}$) of the Gr-MoS$_2$ heterostructure (a-c) and pristine MoS$_2$ (d-f) FETs, measured at drain voltages ranging from 0.05 to 1.0 V over the temperature range of 300-400 K. Figure 4 corresponds to single-sweep (forward sweep) measurements, while Fig. S7 shows dual-sweep (forward and reverse) characteristics, providing insight into the temperature-dependent transport and hysteresis behavior of both devices. For both devices, $I_{ds}$ increases monotonically with positive gate voltage at all temperatures, confirming stable n-type transport. An increase in $I_{ds}$ is observed with increasing temperature, particularly at higher gate and drain biases, indicating thermally assisted carrier injection and a reduction in the effective contact barrier height. In pristine MoS$_2$, this behaviour can be attributed to thermally activated transport through localized states and temperature-assisted carrier injection across the metal-MoS$_2$ interface. This result is consistent with Ahmed et al. who reported carrier transport across the metal-MoS$_2$ interface is governed by thermionic emission at high temperatures, indicating thermally assisted carrier injection over the Schottky barrier.[37] A closer inspection of Fig. 4(a-c) reveals that $I_{ds}$ in the Gr-MoS$_2$ heterostructure increases more rapidly with rising temperature compared to pristine MoS$_2$. This behaviour can be attributed to improved carrier injection facilitated by graphene, along with reduced influence of interfacial trap states and more efficient charge transport at elevated temperatures. This behaviour is consistent with the enhancement in mobility and conductivity observed at RT for the Gr-MoS$_2$ heterostructure and supports the proposed interfacial charge transfer mechanism, indicating that graphene acts as an effective contact layer that facilitates carrier transfer and improves interfacial transport.

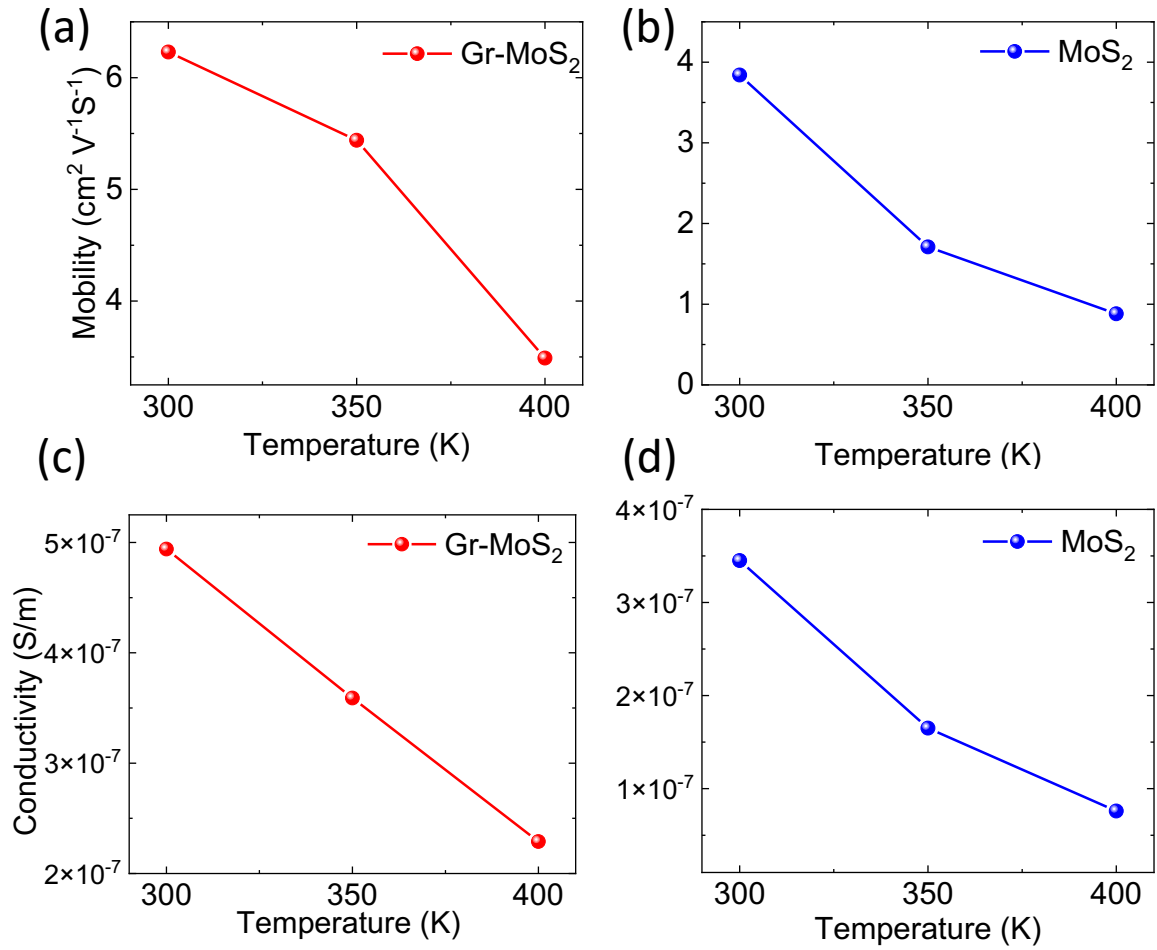

*Figure 5:* *Temperature-dependent electrical transport properties of pristine MoS$_2$ and Gr-MoS$_2$ heterostructure field-effect transistors. Field-effect mobility of (a) Gr-MoS$_2$ heterostructure and (b) pristine MoS$_2$ as a function of temperature. Electrical conductivity of (c) Gr-MoS$_2$ heterostructure device and (d) pristine MoS$_2$ device with temperature.*

**Temperature-Dependent Mobility and Conductivity Analysis:**

The temperature dependence of the field-effect mobility (μ) and electrical conductivity (σ) of Gr-MoS$_2$ heterostructure and pristine MoS$_2$ field-effect transistors are investigated in the range of 300-400 K to elucidate the role of graphene contact engineering on charge transport. At room temperature (300 K), the pristine MoS$_2$ device with Ag electrodes exhibits a mobility of 3.84 cm$^2$V$^{-1}$s$^{-1}$ and a conductivity of 3.45 × 10$^{-7}$ S/m, whereas the Gr-MoS$_2$ heterostructure shows enhanced values of 6.23 cm$^2$V$^{-1}$s$^{-1}$ and 4.94 × 10$^{-7}$ S/m, respectively, corresponding to a mobility enhancement factor [μ(Gr-MoS$_2$)/μ(MoS$_2$)] of ~1.62 and a conductivity enhancement factor [σ(Gr-

MoS$_2$)/σ(MoS$_2$)] of ~1.43. This improvement at RT indicates more efficient carrier injection and reduced contact resistance at the Gr-MoS$_2$ interface compared to conventional metal-semiconductor contacts. Cheng et.al reported that with increasing temperature the field effect mobility of MoS$_2$ based devices decreases due to enhancement in phonon scattering.[38] Consistent with their report, our results show that with increasing temperature, the μ and σ of both devices decrease monotonically, a characteristic of two-dimensional semiconductors where phonon scattering becomes increasingly dominant. The conductivity follows the relation $\sigma = ne\mu$, and its temperature dependence closely mirrors that of the mobility, suggesting that the observed transport behaviour is primarily governed by mobility degradation rather than significant changes in carrier concentration. In the pristine MoS$_2$ device, the mobility decreases sharply to 1.71 cm$^2$V$^{-1}$s$^{-1}$ at 350 K and 0.88 cm$^2$V$^{-1}$s$^{-1}$ at 400 K, representing an overall reduction in mobility of ~77%, accompanied by a corresponding drop in conductivity to 1.65 × 10$^{-7}$ S/m (350K) and 0.76 × 10$^{-7}$ S/m (400K). Such a steep decline suggests that charge transport in the pristine MoS$_2$ device is strongly influenced by contact resistance and Schottky barrier inhomogeneity at the Ag-MoS$_2$ interface, in addition to enhanced phonon scattering, which becomes increasingly pronounced at elevated temperatures. In contrast, the Gr-MoS$_2$ heterostructure retains significantly higher transport parameters across the entire temperature range, with mobility values of 5.44 cm$^2$ V$^{-1}$ s$^{-1}$ at 350 K and 3.49 cm$^2$ V$^{-1}$ s$^{-1}$ at 400 K, and corresponding conductivity values of 3.59 × 10$^{-7}$ S/m and 2.29 × 10$^{-7}$ S/m. The overall reduction in mobility is limited to ~44%, demonstrating improved thermal stability. Consequently, the mobility and conductivity enhancement factors increase markedly with temperature, reaching ~3.18 (mobility) and ~2.18 (conductivity) at 350 K, and ~3.97 and ~3.01, respectively, at 400 K. This trend highlights the effectiveness of graphene contacts in preserving carrier transport under thermal stress. The improved performance of the Gr-MoS$_2$ device can be attributed to the atomically sharp vdW interface, which minimizes interface trap states and suppresses Schottky barrier fluctuations. Furthermore, the tunable work function of graphene enables improved Fermi-level alignment with the MoS$_2$ conduction band, facilitating efficient carrier injection and reducing contact resistance. As a result, transport in the heterostructure moves closer to a phonon-dominated regime, whereas the pristine MoS$_2$ device remains significantly influenced by contact-limited transport. Overall, these results demonstrate that graphene contact engineering not only enhances room-temperature performance but also significantly improves the thermal robustness of $MoS_2$ based field-effect transistors.

**Table I:** Temperature-dependent carrier transport parameters of pristine $MoS_2$ and Gr-$MoS_2$ heterostructure FETs in the 300-400 K range.

| Temperature (K) | μ($MoS_2$) ($cm^2V^{-1}S^{-1}$) | μ(Gr-$MoS_2$) ($cm^2V^{-1}S^{-1}$) | μ(enhancement factor) (Gr-$MoS_2$)/($MoS_2$) | σ($MoS_2$) ×$10^{-7}$ S/m | σ(Gr-$MoS_2$) ×$10^{-7}$ S/m | σ(enhancement factor) (Gr-$MoS_2$)/($MoS_2$) |
|---|---|---|---|---|---|---|
| 300 | 3.84 | 6.23 | 1.62 | 3.45 | 4.94 | 1.43 |
| 350 | 1.71 | 5.44 | 3.18 | 1.65 | 3.59 | 2.18 |
| 400 | 0.88 | 3.49 | 3.97 | 0.76 | 2.29 | 3.01 |

The temperature dependence of the carrier transport is analysed by fitting the experimental data using a power-law relation of the form,

$$\mu, \sigma \propto T^{-\gamma} \qquad (3)$$

The temperature exponent (γ) is extracted from the slopes of the log-log plots of mobility (ln μ) and conductivity (ln σ) as a function of temperature, obtained via linear fitting, as shown in Figure S8 of the Supplementary Information. For the pristine $MoS_2$ device, the mobility and conductivity exhibit strong temperature dependence with slopes of -5.124 and -5.246, respectively, corresponding to $\mu_{MoS_2} \propto T^{-5.124}$ and $\sigma_{MoS_2} \propto T^{-5.246}$. Previous studies have reported that the temperature dependence of mobility in $MoS_2$ FETs typically yields γ values in the range of ~1.5-1.7, consistent with phonon-limited transport. For instance, first-principles calculations for monolayer $MoS_2$ report γ ~ 1.69 for phonon-limited mobility near room temperature.[39] However, in our case, the extracted γ values are significantly higher, suggesting that the transport is not purely phonon-limited. These comparatively large γ values for pristine $MoS_2$ indicate enhanced temperature-dependent scattering, likely arising from the combined influence of contact resistance, trap-assisted transport, and intrinsic phonon scattering mechanisms. In contrast, the Gr-$MoS_2$ heterostructure shows a significantly weaker temperature dependence, with slopes of -1.985 for mobility and -2.657 for conductivity, yielding $\mu_{\text{Gr-MoS}_2} \propto T^{-1.985}$ and $\sigma_{\text{Gr-MoS}_2} \propto T^{-2.657}$. The extracted temperature exponents (γ) obtained from power-law fitting of mobility and conductivity $\mu, \sigma \propto T^{-\gamma}$ are summarized in Table S1 of the Supplementary Information.

A quantitative comparison of the extracted exponents reveals that the temperature sensitivity of carrier transport is substantially reduced in the heterostructure. Specifically, the mobility exponent decreases by a factor of ~2.6 (from 5.124 to 1.985), while the conductivity exponent decreases by ~2.0 (from 5.246 to 2.657), indicating a significantly weaker dependence on temperature in the Gr-MoS$_2$ device. The markedly reduced temperature exponent in the Gr-MoS$_2$ device suggests a reduction in temperature-activated scattering processes and improved carrier transport compared to pristine MoS$_2$. This behaviour can be attributed to the presence of graphene, which facilitates efficient charge transfer, reduces contact resistance, and screens Coulomb impurities at the interface as discussed earlier. Consequently, the heterostructure exhibits enhanced thermal stability and weaker temperature-dependent degradation, highlighting its potential for better-performance electronic applications.

**Conclusion:**

In summary, we have systematically investigated the structural, optical, and temperature-dependent electrical transport properties of graphene-MoS$_2$ van der Waals heterostructure field-effect transistors and compared them with pristine MoS$_2$ devices. Optical microscopy, Raman spectroscopy, photoluminescence quenching, and AFM analysis confirm the formation of a clean few layer graphene/monolayer MoS$_2$ interface with efficient interlayer coupling. Electrical measurements reveal enhanced drain current, improved field-effect mobility, and higher conductivity in the Gr-MoS$_2$ heterostructure relative to conventional Ag-contacted MoS$_2$ devices. Band-alignment considerations suggest that graphene facilitates more efficient carrier injection by reducing effective Schottky barrier effects at the interface. Temperature-dependent studies (300-400 K) demonstrate that although both devices exhibit mobility degradation consistent with phonon-dominated transport, the heterostructure maintains significantly higher mobility and conductivity with comparatively reduced thermal degradation. The mobility and conductivity enhancement factors increase markedly with temperature, highlighting the superior thermal stability of graphene-engineered contacts. The strong correlation between mobility and conductivity trends suggests that transport enhancement is primarily associated with improved charge injection rather than significant changes in carrier concentration. The mobility enhancement factor increases from ~1.6 at 300 K to ~4.0 at 400 K, accompanied by a corresponding improvement in conductivity stability, demonstrating superior thermal robustness. Power-law fitting ($\mu, \sigma \propto T^{-\gamma}$) yields exponents of $\gamma \sim 5.124$ and 5.246 for mobility and conductivity in pristine $MoS_2$, whereas reduced values of $\gamma \sim 1.985$ and 2.657 are observed for the Gr-$MoS_2$ device. This corresponds to a reduction factor of ~2.6 for mobility and ~2.0 for conductivity indicating reduced temperature sensitivity and improved carrier transport stability. Overall, this work demonstrates that interfacial charge-transfer engineering using graphene contacts as an effective strategy toward achieving transport behaviour closer to intrinsic limits in $MoS_2$ transistors, offering a robust pathway toward thermally stable, better-performance 2D electronic devices.

**Associated Contents**

**Supporting information:**

Optical images of monolayer (ML) $MoS_2$, few-layer (FL) graphene, and the FL Gr-ML $MoS_2$ heterostructure, Raman intensity mapping of the Gr-$MoS_2$ heterostructure, individual Raman spectra of FL graphene and ML $MoS_2$, atomic force microscopy (AFM) images of the Gr-$MoS_2$ heterostructure, optical image of a fabricated device on pristine ML $MoS_2$, room-temperature and temperature-dependent transfer characteristics of Gr-$MoS_2$ heterostructure and pristine ML $MoS_2$ devices, log-log plots of mobility (*ln μ*) and conductivity (*ln σ*) as a function of temperature for both devices, and a table summarizing the extracted temperature exponents ($\gamma$) from power-law fitting ($\mu, \sigma \propto T^{-\gamma}$).

**Acknowledgments**

S.P.S. and P.S. acknowledge Anusandhan National Research Foundation (ANRF) (CRG/2023/006935) for partial financial support and fellowship, respectively. G.K.P. acknowledges the use of the micro-Raman facility at the Central Research Facility (CRF) of KIIT Deemed to be University, Bhubaneswar.